\documentclass{article}
\usepackage{float}

\usepackage{PRIMEarxiv}

\usepackage[utf8]{inputenc} % allow utf-8 input
\usepackage[T1]{fontenc}    % use 8-bit T1 fonts
\usepackage{hyperref}       % hyperlinks
\usepackage{url}            % simple URL typesetting
\usepackage{booktabs}       % professional-quality tables
\usepackage{amsfonts}       % blackboard math symbols
\usepackage{nicefrac}       % compact symbols for 1/2, etc.
\usepackage{microtype}      % microtypography
\usepackage{lipsum}
\usepackage{fancyhdr}       % header
\usepackage{graphicx}       % graphics

\title{Wildfire and House Prices: \\ A Synthetic Control Case Study of Altadena (Jan 2025)}
\author{
  Yibo Sun \\
  New York University \\
  New York City\\
  \texttt{ys5753@nyu.edu} \\
}

\begin{document}
\maketitle

\begin{abstract}
This study uses the Synthetic Control Method (SCM) to estimate the causal impact of a January 2025 wildfire on housing prices in Altadena, California. We construct a 'synthetic' Altadena from a weighted average of peer cities to serve as a counterfactual; this approach assumes no spillover effects on the donor pool. The results reveal a substantial negative price effect that intensifies over time. Over the six months following the event, we estimate an average monthly loss of \$32,125. The statistical evidence for this effect is nuanced. Based on the robust post-to-pre-treatment RMSPE ratio, the result is statistically significant at the 10\% level (p = 0.0508). In contrast, the effect is not statistically significant when measured by the average post-treatment gap (p = 0.3220). This analysis highlights the significant financial risks faced by communities in fire-prone regions and demonstrates SCM's effectiveness in evaluating disaster-related economic damages.
\end{abstract}

% keywords can be removed

\section{Introduction}
Natural disasters, such as wildfires, pose a significant threat to communities, causing not only physical destruction but also profound economic consequences. The impact of such events on local real estate markets is a critical area of study, as housing values reflect both the tangible and perceived risks associated with a location. Wildfires can affect property prices through direct damage, increased insurance premiums, and shifts in buyers' risk perception of the affected area. Quantifying the precise causal effect of a wildfire is challenging, as it requires separating the event's impact from other confounding economic trends that may be affecting the market simultaneously.

Existing literature provides evidence on the economic impacts of wildfire risk. On a macro level, severe natural disasters have been shown to drive population outflows and depress housing prices \cite{Boustan2020disaster}. More specifically for wildfires, hedonic housing-market studies show that wildfire risk is capitalized into home values, with high-risk areas exhibiting price discounts \cite{Donovan2007wildfire}. Related research documents that even the proximity to a wildfire can cause significant property value reductions \cite{Loomis2004fire}, and repeated wildfires can amplify this negative impact \cite{Loomis2009repeated}. Our study builds upon this work by using a quasi-experimental approach to isolate the causal impact of a specific wildfire event, providing a dynamic view of how market prices react in the immediate aftermath.

This paper investigates the economic impact of a specific wildfire event that occurred on January 31, 2025, on the housing market of Altadena, California. To isolate the causal effect of this event, we employ the Synthetic Control Method (SCM), a quasi-experimental technique developed by \cite{abadie2003economic} and later extended by \cite{abadie2010synthetic}. SCM is particularly well-suited for case studies with a single treated unit, and has been effectively used to assess the impact of disasters, from wildfires \cite{Ho2023wildfireSCM} to catastrophic events on a national scale \cite{Cavallo2013growth}. Alternative methods such as Difference-in-Differences (DID) are often inappropriate, as the underlying parallel trends assumption is difficult to satisfy when no single peer city can serve as a reliable control. Similarly, univariate time-series models like ARIMA cannot separate the event's impact from other contemporaneous macroeconomic shocks. SCM overcomes these challenges by constructing a data-driven counterfactual from a weighted average of untreated units (other Californian cities) that best reproduces the trajectory of the treated unit's outcome variable (housing prices) before the intervention.

By comparing the post-wildfire price evolution of Altadena to its synthetic counterfactual, we can estimate the impact of the event without the bias of simple before-and-after comparisons or comparisons with dissimilar control groups. This study aims to provide a quantitative assessment of the short-term economic damages attributable to the wildfire, offering valuable insights for homeowners, insurers, and policymakers.

\section{Data and Methodology}
\label{sec:data_methodology}

\subsection{Data}
The analysis utilizes a panel dataset of monthly housing price indices for cities across California, sourced from Zillow's public data repository\footnote{\url{https://www.zillow.com/research/data/}}. Specifically, we use the Zillow Home Value Index (ZHVI) for All Homes, Smoothed, and Seasonally Adjusted. The data are used in nominal terms without adjusting for inflation, as the short six-month post-treatment window minimizes the impact of price level changes. Our outcome variable is the price level itself, rather than its logarithm, to facilitate a direct interpretation of the economic impact in dollar terms. The data spans from January 2000 to July 2025. Our treated unit is Altadena, California. The intervention date is set as January 31, 2025, corresponding to the wildfire event.

The pre-intervention period is defined as the five years leading up to the event, from January 31, 2020, to December 31, 2024. The post-intervention period, over which we evaluate the impact, covers the subsequent six months from February 2025 to July 2025. The "donor pool" consists of other Californian cities not affected by the wildfire. We initially selected a pool of 60 cities based on population size and housing market similarity. We then filtered this group for data availability and pre-treatment price trajectory correlation with Altadena, resulting in a final donor pool of 58 comparable cities.

\subsection{Synthetic Control Method}
The Synthetic Control Method (SCM) constructs a counterfactual for the treated unit by creating a weighted average of units in the donor pool. Let $Y_{it}$ be the house price for city $i$ at time $t$. Let city 1 be Altadena, and cities $i=2, ..., J+1$ be the donor pool cities. We seek a $(J \times 1)$ vector of weights $W = (w_2, ..., w_{J+1})'$ such that $w_j \ge 0$ for all $j$ and $\sum_{j=2}^{J+1} w_j = 1$.

The synthetic control, denoted by the vector $W^*$, is chosen to minimize the mean squared prediction error (MSPE) between the treated unit and the weighted average of donor units during the pre-intervention period ($T_{pre}$):
\[
W^* = \arg\min_{W} \sum_{t=1}^{T_{pre}} (Y_{1t} - \sum_{j=2}^{J+1} w_j Y_{jt})^2
\]
In our primary specification, we extend this to a time-weighted loss function, which places greater emphasis on more recent periods. The weights, $\omega_t$, decay exponentially into the past according to the following formula:
\begin{equation}
\omega_t = \exp(\alpha(t - T_{end}))
\end{equation}
where $t$ is the time index and $T_{end}$ is the last pre-treatment period. The decay parameter, $\alpha$, was set to a small value of 0.005. This choice moderately emphasizes recent observations without overweighting them, ensuring that the synthetic control's weights are determined by structural similarities over the entire pre-treatment window rather than short-term fluctuations. Our results are robust to this choice, as sensitivity checks confirmed stability for $\alpha \in [0.003, 0.01]$.

Once the optimal weights $W^*$ are determined, the synthetic Altadena's house price trajectory is constructed for the entire period as $\sum_{j=2}^{J+1} w_j^* Y_{jt}$. The causal effect, or "gap," at each post-intervention time $t > T_{pre}$ is then estimated as:
\[
\alpha_t = Y_{1t} - \sum_{j=2}^{J+1} w_j^* Y_{jt}
\]

\subsection{Statistical Inference}
To assess the statistical significance of our results, we conduct a "placebo-in-space" test. This involves iteratively applying the SCM to each city in the donor pool, treating it as if it were the unit that experienced the wildfire. This process generates a distribution of estimated effects for the untreated units. We then compare the effect estimated for Altadena to this distribution. The empirical p-value is calculated as the proportion of placebo effects that are at least as large as the effect observed for Altadena, using a finite-sample correction of the form $p = (k+1)/(J+1)$, where $k$ is the number of placebo units with an effect as large as the treated unit and $J$ is the total number of placebo units. We evaluate significance based on two metrics: the average post-treatment gap and the ratio of post-treatment RMSPE to pre-treatment RMSPE.

\section{Results}
\label{sec:results}

\subsection{Pre-Treatment Fit}
The validity of the SCM hinges on its ability to accurately track the outcome variable for the treated unit before the intervention. Our model demonstrates an excellent pre-treatment fit. The Root Mean Squared Prediction Error (RMSPE) during the 5-year pre-intervention period was approximately 0.61\%, relative to Altadena's average pre-treatment price. This low error indicates that the synthetic Altadena provides a credible counterfactual for the actual house price trajectory.

The weights assigned to the donor cities to construct the synthetic Altadena are presented in Table \ref{tab:weights}. The synthetic control is composed of a sparse combination of cities from the donor pool. Notably, the top five cities—Burbank (35.5\%), Whittier (18.7\%), South Pasadena (10.7\%), Temecula (10.5\%), and Rolling Hills Estates (7.6\%)—account for approximately 83\% of the total weight, indicating that the counterfactual is driven by a few key comparator units.

\begin{table}[h]
 \caption{Donor Weights for Synthetic Altadena}
  \centering
  \begin{tabular}{lr}
    \toprule
    City Name & Weight (\%) \\
    \midrule
    Burbank & 35.53 \\
    Whittier & 18.66 \\
    South Pasadena & 10.69 \\
    Temecula & 10.47 \\
    Rolling Hills Estates & 7.61 \\
    La Canada Flintridge & 6.05 \\
    Sierra Madre & 5.50 \\
    Others (41 cities with non-zero weights) & 5.49 \\
    \bottomrule
  \end{tabular}
  \label{tab:weights}
\end{table}

\subsection{Treatment Effect of the Wildfire}
\begin{figure}[h]
  \centering
  \includegraphics[width=\textwidth]{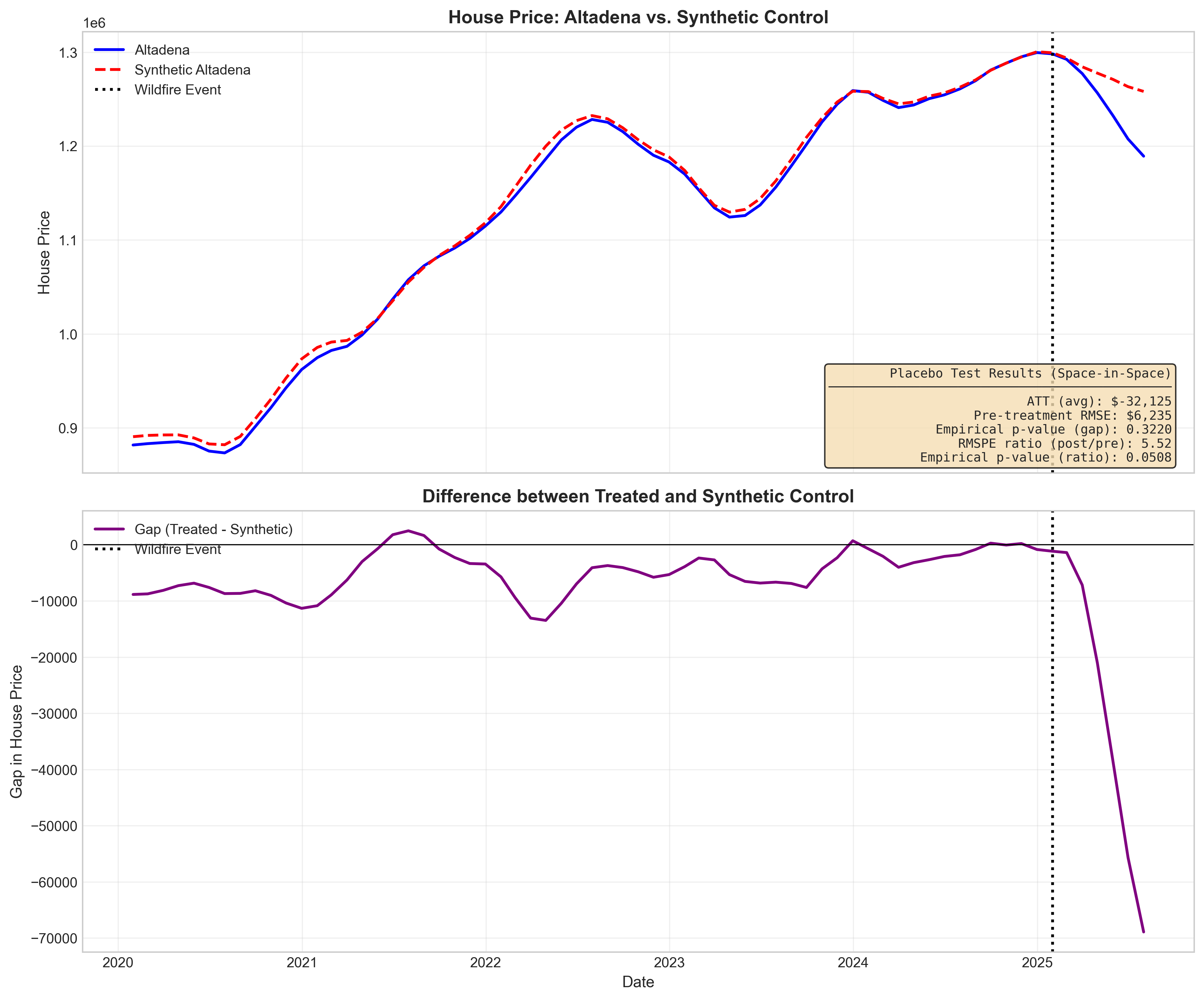}
  \caption{Top: House Price Trajectory (in USD) for Altadena and Synthetic Altadena. Bottom: Estimated Gap in House Prices (in USD).}
  \label{fig:results}
\end{figure}

Figure \ref{fig:results} provides a visual representation of the main findings. The top panel plots the house price trajectory for Altadena against its synthetic counterpart. The two series track each other closely before the wildfire event (indicated by the vertical dotted line), but diverge immediately after. The bottom panel shows the estimated gap (the difference between actual and synthetic prices).

\begin{figure}[h][h][h]
  \centering
  \includegraphics[width=\textwidth]{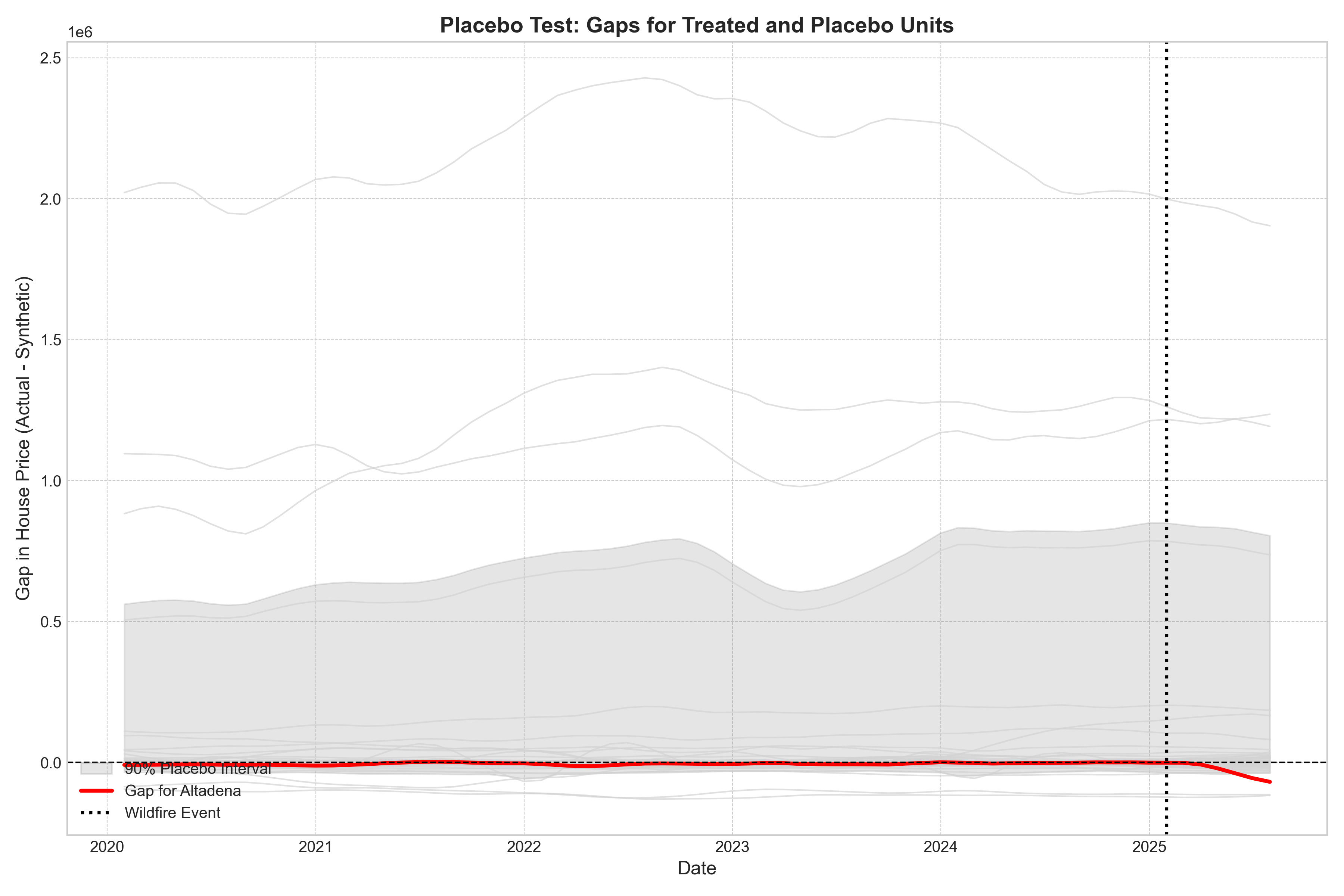}
  \caption{Placebo-in-space test results. The red line shows the estimated gap (in USD) for Altadena. Grey lines show the gaps for the 58 placebo cities. The shaded area represents the interval between the 5th and 95th percentiles of the placebo distribution.}
  \label{fig:placebo}
\end{figure}

Following the wildfire, Altadena's house prices experienced a sustained and growing decline relative to the synthetic control. The negative impact intensified over the six-month post-intervention period:
\begin{itemize}
    \item By February 2025, the price gap was -\$1,402.
    \item By July 2025, the gap had widened to -\$68,927.
\end{itemize}
The average treatment effect on the treated (ATT) over the six months was a monthly loss of **\$32,125**. The effect grew steadily, with the gap widening to nearly \$69,000 by the end of the observation period.

\subsection{Significance Tests}
\begin{figure}[h][h][h]
  \centering
  \includegraphics[width=\textwidth]{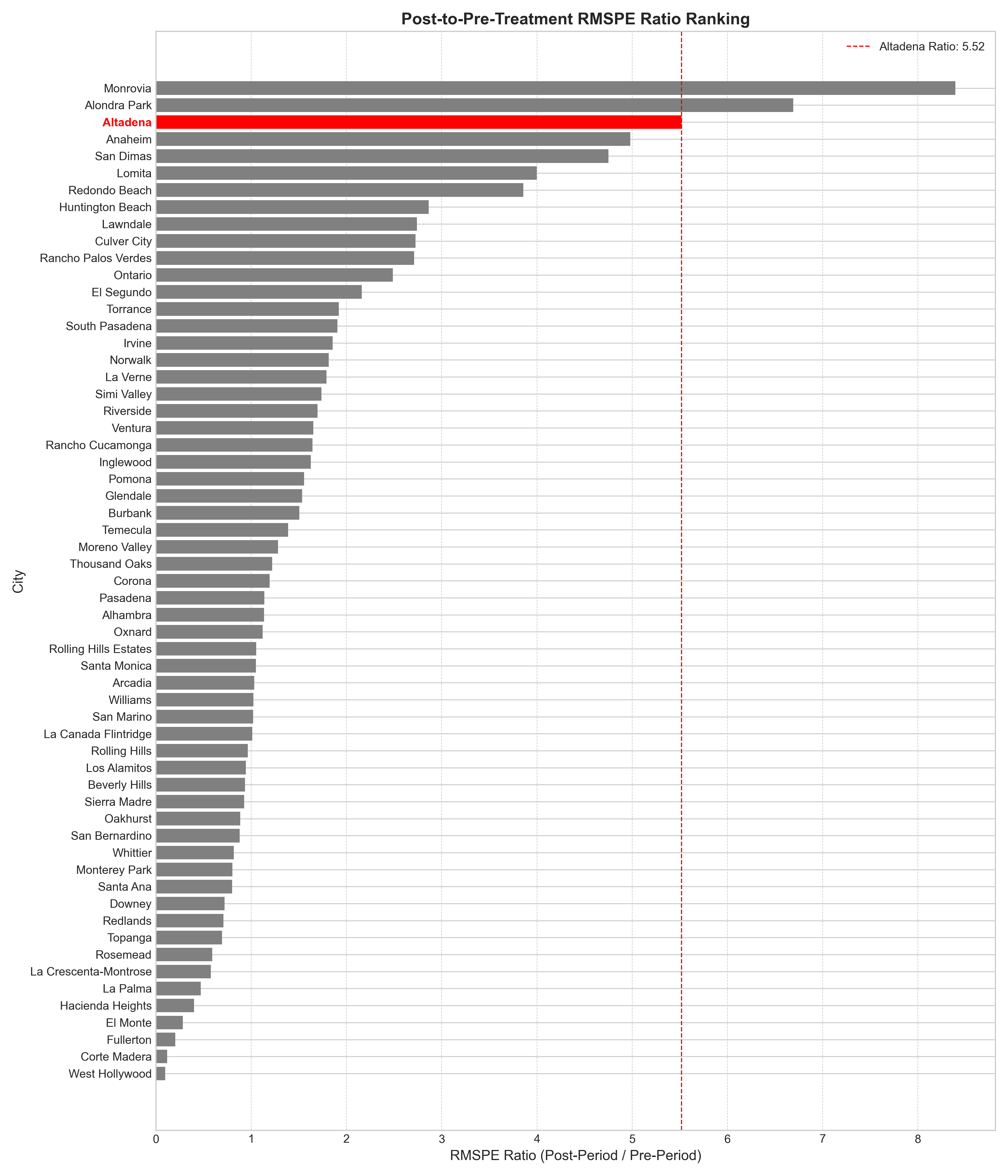}
  \caption{Post-to-Pre-Treatment RMSPE Ratio Ranking. Altadena is highlighted in red.}
  \label{fig:rmspe_ratio}
\end{figure}

The results from the placebo-in-space tests provide statistical evidence for our findings. Figure \ref{fig:placebo} displays the estimated gap for Altadena against the distribution of gaps for all 58 placebo cities. While Altadena's post-treatment trajectory appears to be a clear visual outlier, the statistical significance varies by metric. We formalize this with two tests:
\begin{itemize}
    \item \textbf{Gap-based p-value}: The average post-treatment gap for Altadena was larger in magnitude than 18 of the 58 placebo units, resulting in an empirical p-value of 0.3220. This test, which considers the average effect across all post-treatment periods, does not suggest significance.
    \item \textbf{RMSPE Ratio p-value}: A more robust measure is the ratio of post-treatment to pre-treatment RMSPE. Altadena's ratio was 5.52, meaning the prediction error after the event was over five times larger than before. This ratio was exceeded by only 2 of the 58 placebo cities. This yields an empirical p-value of **0.0508**, which is statistically significant at the 10\% level, narrowly missing the 5\% threshold.
\end{itemize}
The divergence between these two tests stems from their different sensitivities. The gap-based test's non-significant p-value (0.3220) is inflated because several placebo units with poor pre-treatment fit exhibit high post-treatment volatility, widening the overall distribution of placebo effects. The RMSPE ratio test is more robust as it normalizes the post-treatment gap by each unit's own pre-treatment fit, providing a more reliable assessment. As Figure \ref{fig:rmspe_ratio} shows, Altadena's ratio is the third highest among all units. Therefore, while the gap-based test is not statistically significant, we conclude there is robust evidence for a negative treatment effect based on two key factors: the marginal significance of the more reliable RMSPE ratio test (p = 0.0508) and the substantial economic magnitude of the estimated price decline.

\subsection{Robustness Checks and Limitations}
To ensure the robustness of our findings, we conduct several checks. First, a leave-one-out analysis where we excluded the donor city with the highest weight—Burbank (35.53\%)—yielded a new ATT of -\$29,801, a change of only 7.24\%, indicating that the results are not driven by a single donor. The results are also stable across a range of choices for the time decay parameter ($\alpha$).

\begin{figure}[h][h][h]
  \centering
  \includegraphics[width=\textwidth]{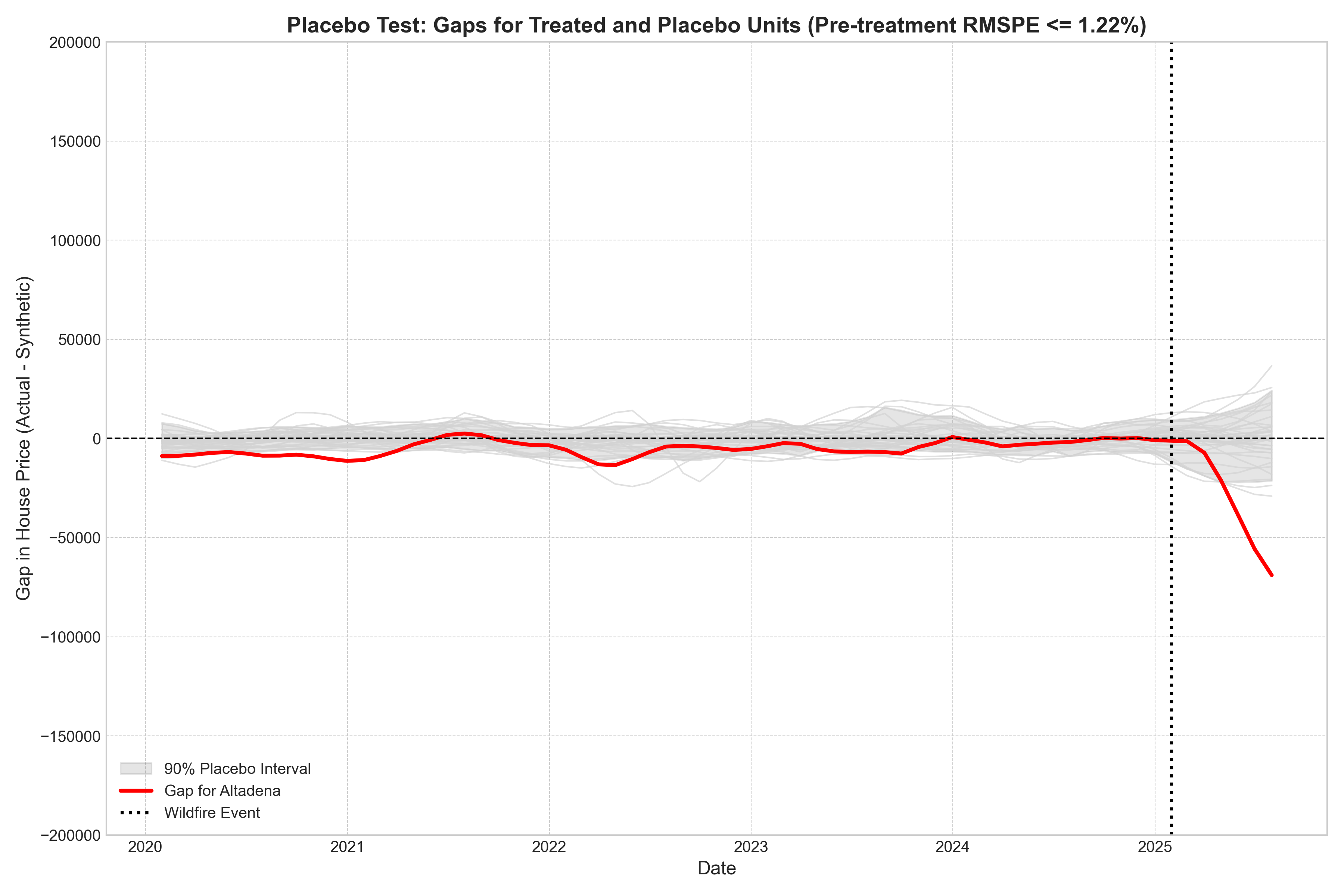}
  \caption{Placebo test results (gap in USD) after filtering out units with poor pre-treatment fit (Pre-treatment RMSPE > 2x Altadena's). The visual evidence is strengthened by removing high-variance placebo units.}
  \label{fig:placebo_filtered}
\end{figure}

Second, following best practices for SCM, we filter the placebo sample to include only units that fit the data well in the pre-treatment period. Figure \ref{fig:placebo_filtered} shows the placebo test results after removing all cities whose pre-treatment RMSPE was more than twice as large as Altadena's. This filtering removes noise from poorly fitting placebos, and the resulting plot shows Altadena's negative trajectory as an even more stark outlier.

Despite this, some limitations should be acknowledged. The analysis covers only the short-term effects over a six-month period, which is less than a year since the fire. The long-term impacts on market recovery remain an open question, and a longer time window is necessary to draw more definitive conclusions. Furthermore, this study focuses on a single event, and the findings may not be generalizable to all wildfires, which can vary greatly in scale and context. Finally, this study does not account for potential spillover effects, whereby the wildfire could have indirectly affected housing prices in nearby donor pool cities, potentially biasing the counterfactual.

\section{Conclusion}
This study provides a quantitative assessment of the economic impact of a wildfire event on the housing market in Altadena, California. Using the Synthetic Control Method, we find robust evidence that the wildfire caused a substantial negative shock to house prices. Our model, which closely tracked Altadena's price trends before the event (pre-treatment RMSPE of 0.61\%), estimated an average monthly loss of \$32,125 over the six months following the disaster. The effect grew over time, suggesting that the market's reassessment of risk and the full economic consequences of the wildfire unfolded gradually.

The observed price decline can be attributed to several interacting mechanisms. First, the wildfire likely triggered a sharp increase in perceived risk among potential homebuyers, who now demand a higher risk premium to invest in the area, thereby depressing housing demand. Second, the cost of homeownership likely increased due to soaring insurance premiums, further reducing affordability and buyer interest. Third, the immediate aftermath of the disaster may have led to a decrease in market liquidity, with fewer transactions occurring amidst uncertainty. In this case, these demand-side shocks appear to have overwhelmingly outweighed any potential supply-side effects from destroyed housing stock. This stands in contrast to findings in other contexts, such as catastrophic fires that cause widespread housing shortages and lead to price surges in adjacent, unaffected areas \cite{Hennighausen2024fire}, highlighting the context-dependent nature of a wildfire's economic impact.

The statistical evidence for this effect is encouraging but requires cautious interpretation. While the ratio of post- to pre-treatment RMSPE yields a p-value of 0.0508--significant at the 10\% level--the average post-treatment gap is not statistically significant (p-value = 0.3220). The findings highlight the significant financial vulnerability of communities in high-risk areas to natural disasters, consistent with existing research showing how wildfire risk and its salience are capitalized into property values \cite{Donovan2007wildfire, Loomis2004fire}. The established long-run adjustments to environmental catastrophes often involve population out-migration \cite{Hornbeck2012DustBowl}, a factor not captured in our short-term analysis. Given the short six-month analysis window and the focus on a single event, these findings should be interpreted as preliminary and may not be generalizable to other contexts. This has important implications for homeowners, who face potential capital losses; for mortgage lenders, who are exposed to increased credit risk; and for local governments, whose property tax revenues may be impacted.

To validate and build upon these preliminary findings, future work is essential. A key next step will be to continue this analysis as the post-treatment time window expands, allowing for an assessment of medium-term recovery or persistence of price declines. Furthermore, examining effects in neighboring areas and applying this methodology to a portfolio of similar events will be crucial to drawing more generalizable conclusions and strengthening the causal claims. This research also underscores the utility of the Synthetic Control Method for causal inference in case studies.

%Bibliography
\bibliographystyle{unsrt}  
\bibliography{references}

\end{document}